\documentclass[a4paper,11pt]{article}
\usepackage{pos}

\usepackage{natbib}
\title{Implications of Mini-EUSO measurements for a space-based observation of UHECRs}
 \ShortTitle{Implications of Mini-EUSO measurements}

\author*[a,b]{Mario Bertaina}
\author[b,i]{Matteo Battisti}
\author[a,b]{Marta Bianciotto}
\author[j]{Karl Bolmgren}
\author[p,m]{Francesco Fenu}
\affiliation[a]{University of Turin, Department of Physics, V. P. Giuria 1, 10125 Turin, Italy}
\affiliation[b]{INFN Section of Turin, Via P. Giuria 1, 10125 Turin, Italy}
\affiliation[i]{Univ. Paris Cit\'e,CNRS, Astroparticule \& Cosmologie,10 Rue A. Domon \& L. Duquet, 75013 Paris,France}
\affiliation[j]{KTH Royal Institute of Technology, Brinellv\"gen 8, 114 28 Stockholm, Sweden}
\affiliation[m]{ASI, Italian Space Agency, Via del Politecnico, 00133 Rome, Italy}
\affiliation[p]{KIT, Hermann-von-Helmholtz-Platz 1, 76344 Eggenstein-Leopoldshafen, Germany}
\onbehalf{for the JEM-EUSO Collaboration\\[-1mm]{\normalsize \normalfont (a complete list of authors can be found at the end of the proceedings)}}
\emailAdd{bertaina@to.infn.it}
\abstract{Mini-EUSO is the first mission of the JEM-EUSO program on board the International Space Station. It was launched in August 2019 and it is operating since October 2019 being
located in the Russian section (Zvezda module) of the station and viewing our planet from a nadir-facing UV-transparent window. The instrument is based 
on the concept of the original JEM-EUSO mission and consists of an optical system employing two Fresnel lenses of 25 cm each and a focal surface composed of 36 
Multi-Anode Photomultiplier tubes, 64 channels each, for a total of 2304 channels with single photon counting sensitivity and an overall field of view
of 44$\times$44$^\circ$. Mini-EUSO can map the night-time Earth in the near UV range (predominantly between 290 nm and 430 nm), with a spatial
resolution of about 6.3 km and different temporal resolutions of 2.5 $\mu$s, 320 $\mu$s and 41 ms. Mini-EUSO observations are extremely important to
better assess the potential of a space-based detector in studying Ultra-High Energy Cosmic Rays (UHECRs) such as K-EUSO and POEMMA. In this contribution
we focus the attention on the results of the UV measurements and we place them in the context of UHECR observations from space, namely the estimation of exposure.
}

\ConferenceLogo{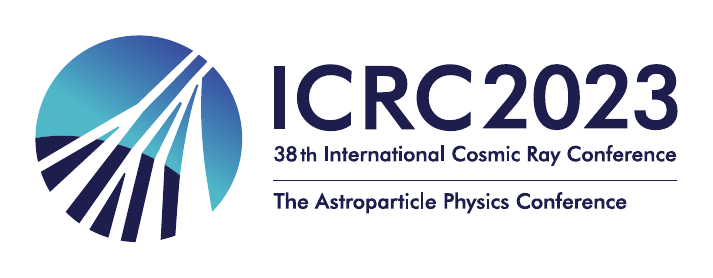}

\FullConference{%
38th International Cosmic Ray Conference (ICRC2023)\\
  26 July - 3 August, 2023\\
  Nagoya, Japan}

\begin{document}
\maketitle

\section{Introduction}
\label{sec:intro}
The current main goal in the field of UHECR (Ultra-High Energy Cosmic Ray) science is to identify their
astrophysical sources and composition~\cite{snowmass}.
For this, increased statistics is one of the essential requirements. A space-based detector for
UHECR research has the advantage of a very large exposure and a uniform coverage of the celestial sphere.
The aim of the JEM-EUSO program~\cite{EUSO-Program2} is to bring the study of UHECRs to space.
The principle of observation is based on the detection of UV light emitted by isotropic ﬂuorescence
of atmospheric nitrogen excited by Extensive Air Showers (EASs) in Earth’s atmosphere and
forward-beamed Cherenkov radiation reﬂected at the Earth’s surface or at dense cloud tops.
%In addition to the prime objective of UHECR studies, a space-based detector will do several secondary
%studies due to the instrument's unique capacity of detecting very weak UV signals with extreme
%time-resolution around 1 $\mu$s: meteors, Transient Luminous Events (TLE), bioluminescence, maps of human
%generated UV light, searches for Strange Quark Matter (SQM) and high-energy neutrinos, and more.
The JEM-EUSO program includes missions on ground (EUSO-TA~\cite{eusota}), on stratospheric
balloons (EUSO-Balloon~\cite{eusobal}, EUSO-SPB1~\cite{spb1}, EUSO-SPB2~\cite{spb2}),
and from space (TUS~\cite{tus}, Mini-EUSO~\cite{minieuso,minieuso2}) employing fluorescence
detectors to demonstrate the feasibility of the UHECR observation from space and
prepare for the large size missions K-EUSO~\cite{keuso} and POEMMA~\cite{poemma}.
Mini-EUSO
%(Multiwavelength Imaging New Instrument for the Extreme Universe Space Observatory, known as
%\emph{UV atmosphere} in the Russian Space Program)~\cite{minieuso}
is the first detector of the JEM-EUSO program to observe the Earth from the International Space
Station (ISS) and to validate from there the observational principle of a space-based detector for UHECR
measurements.

\section{The Mini-EUSO space telescope} 
\label{sec:minieuso}
Mini--EUSO (Multiwavelength Imaging New Instrument for the Extreme Universe Space Observatory, known as
\emph{UV atmosphere} in the Russian Space Program)~\cite{minieuso} is operating on the ISS since October 7th, 2019, and till
June 2023, more than 80 sessions have been performed lasting 12 hours each.
The pouches containing all stored data are then returned to Earth every $\sim$12 months by the Soyuz
spacecraft.
%is a telescope operating in the near UV  
%range, predominantly between 290~nm and 430~nm, with a square focal surface corresponding to a field of view
%(FoV) of $\sim$44$^{\circ}$ $\times$ 44$^{\circ}$. Its spatial resolution at ground level is approximately 
%$6.3 \times 6.3$~km$^2$, slightly varying with the altitude of the ISS and the pointing direction of the
%pixel. 
%The detector size is $37 \times 37 \times 62$~cm$^3$, mainly constrained by the size of the 
%nadir-facing UV transparent window in the Russian Zvezda module, where it is attached 
%a couple of times per month during onboard
%night-time, approximately at 18:30 UTC, with operations lasting about 12 hours.
%The first observations took place on October 7, 2019. 
%Since October 7th, 2019 and till
%June 2023, more than 80 sessions have been performed lasting 12 hours each. 
%Data are stored locally on 512~GB USB pendrives.
%After each data-taking session samples of data (about 10\% of stored data, usually corresponding to the
%beginning and the end of each session) are copied and transmitted to ground to verify the correct
%functioning of the instrument and subsequently optimize its working parameters.

The optics are based on two 25~cm diameter Fresnel lenses in Polymethyl methacrylate (PMMA).
%This material allows for a light (11~mm thickness, 0.87~kg/lens), robust, and compact design well suited for
%space applications.
The Mini--EUSO focal surface, or Photo Detector Module (PDM), consists of a grid of 36 Multi-Anode
Photomultiplier Tubes (MAPMTs, Hamamatsu Photonics  R11265-M64).
%arranged in an array of $6 \times 6$ elements.
Each MAPMT consists of $8 \times 8$ pixels, resulting in a total of 2304 channels.
The MAPMTs are grouped in Elementary Cells (ECs) of $2 \times 2$ MAPMTs. MAPMTs are separated by 2 -- 3 mm
spacing to avoid vibration damage at launch. Consequently, gaps exist between MAPMTs in the collected images.
Each EC has an independent high voltage power supply (HVPS) and board connecting the dynodes and anodes of
the four photomultipliers.
The HVPS system is based on a Cockroft-Walton circuit. The system has an internal safety
mechanism which operates either reducing the collection efficiency of the four MAPMTs or reducing the
MAPMT gain when particularly bright signals occur (i.e. lightning or large cities)~\cite{minieuso}.
%These statuses of reduced efficiency are called {\it cathode-2} mode. The nominal
%working condition is instead called {\it cathode-3} mode (which is the one assumed in the rest of the paper
%unless stated otherwise).
%The switching from {\it cathode-3} to {\it cathode-2} mode are usually due to lightning strikes or due to
%very bright light sources like large cities.
The recovery to the nominal mode takes place only few
ms after the light level has decreased to a sufficiently low value. In the following we will refer to {\it Cathode3} 
for the nominal mode and {\it Cathode2} for the reduced efficiency/gain mode.
%to avoid continuous oscillation between
%{\it cathode-2} and {\it cathode-3} modes when the light level is close to the switching value.
%The effective focal length of the system is 300~mm, with a Point Spread Function (PSF) of 1.2 MAPMT pixels.
%UV bandpass filters (2~mm of BG3 material) with anti-reflective coating are glued in front of the MAPMTs to
%predominantly select wavelengths between 290~nm and 430~nm.

\begin{figure}[h]
\centering
\includegraphics[width=\columnwidth]{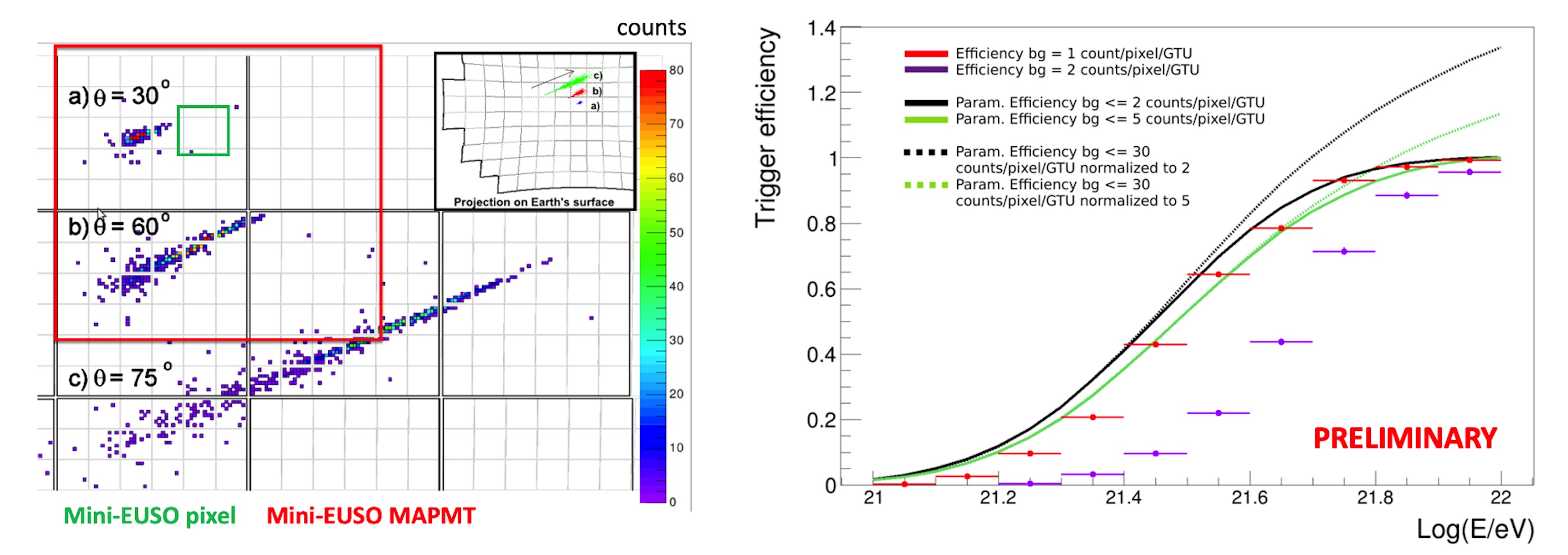}
\caption{Left: Comparison between JEM-EUSO and Mini-EUSO FoVs. The image shows the detected light track of three
simulated EAS with ESAF~\cite{esaf} of energy 10$^{20}$ eV with different zenith angles
($\theta$ = 30$^\circ$, 60$^\circ$, and 75$^\circ$)
as imaged by JEM-EUSO detector. Each little colored dot represents the pixel FoV of JEM-EUSO
projected on ground, each grey square indicates the MAPMT FoV and the large black squares show the FoV of
JEM-EUSO PDMs.
The green and red squares represent Mini-EUSO pixel's and MAPMT's FoV, respectively. Mini-EUSO pixel's FoV
is larger than the FoV of a JEM-EUSO MAPMT and Mini-EUSO MAPMT's FoV is larger than the FoV of one JEM-EUSO PDM.
Right: Trigger efficiency curves of Mini-EUSO for ESAF simulated proton-generated EASs of different
energies on different background levels.
The red and violet points assume a fixed nightglow background of 1 and 2 counts/pix/GTU, respectively.
The continuous black (green) line represents the convolved trigger efficiency curve in which each background level below 2 (5)
counts/pix/GTU is weighted for the relative fraction of time in which it was measured by Mini-EUSO.
The dotted black (green) line provides the fractional increase in exposure [relative to the black (green) line] if
the accepted nightglow background is increased from 2 (5) counts/pix/GTU to 30 counts/pix/GTU. 
A significant increase in exposure is obtained only at the highest energies. See text for details. Image adapted from~\cite{mario-minieuso}.}
\label{fig:fov}       % Give a unique label
\end{figure}

The system has a single photon-counting capability with a double pulse resolution of $\sim$6~ns. Photon
counts are summed in Gate Time Units (GTUs) of 2.5 $\mu$s.
The PDM Data Processor (PDM-DP) stores the 2.5 $\mu$s GTU data stream (D1) in a running buffer on which
the trigger code is executed. 
A detailed description of the trigger algorithm is reported in~\cite{belov-trigger}, which represents
an adaptation of the trigger logic conceived for JEM-EUSO~\cite{mario-trigger}, while the on-board
performance of the trigger system is summarized in~\cite{matteo-trigger}.
The algorithm searches for a signal integrated over 8 consecutive GTUs above 16
standard deviations from the average in any pixel of the focal surface. 
%Both the average and standard
%deviation are calculated in real time to
%take into account varying illumination conditions. In case of a trigger, the 128 frame buffer (64 frames
%before the trigger and 64 after it) is stored in memory.
The choice to operate on 8 GTUs integration and to act on each pixel independently is based on the
time needed by a light signal to cross the FoV of a pixel and the fact that 20 $\mu$s represent a significant
portion of an EAS light-track in atmosphere (see Fig.~\ref{fig:fov} left).

Independently from the trigger, sums of 128 frames (320~\textmu s, D2) are continuously calculated and
stored in another buffer where a similar trigger algorithm, but at this time scale, is running.
Similarly, sums of 128 D2 frames (40.96~ms, D3) are calculated in real time and continuously stored.
Every 5.24~s, 128 GTUs of D3 data, up to 4 D2 packets and up to 4 D1 packets (if triggers were present)
are sent to the CPU for storage. In this way various classes of phenomena with different duration can be
detected with an appropriate time scale (see sec.~\ref{sec:design} for details).

Prior to the launch, the instrument underwent a series of integration and acceptance tests in Rome, Moscow,
and Baikonur, where it was placed in the uncrewed Soyuz capsule. A systematic test of the
acquisition logic was performed at the TurLab facility~\cite{turlab} of the University and at the
Astrophysical Observatory (INAF-OATo)~\cite{Bisconti2021} in Turin.
After launch, an end-to-end in-flight calibration of the Mini-EUSO detector has been performed by assembling
different UV-flasher systems on ground in Japan, Italy, and France and by firing them in various observational
campaigns. A detailed description of the methodology and of the observational results is reported
in~\cite{hiroko2}.

\section{Mini-EUSO design, expected performance and exposure studies}
\label{sec:design}
Mini-EUSO has been designed to detect a photon rate per pixel from diffuse sources (nightglow, clouds, cities, etc.) in the
range of values expected from a large mission in space such as the original JEM-EUSO mission~\cite{jemeuso}
or the future detectors K-EUSO or POEMMA. The pixel FoV is, therefore, $\sim$100 times
larger in area with respect to the FoV of a JEM-EUSO
pixel ($\sim$0.5 $\times$ 0.5 km$^2$), to compensate for the optical system $\sim$100 times smaller,
constrained by the dimension of the UV transparent window (see Fig.~\ref{fig:fov} left).

In order to have a precise ratio of the photon rate per pixel from diffuse sources between JEM-EUSO and
Mini-EUSO, a full simulation of JEM-EUSO and Mini-EUSO detectors was performed with ESAF simulation
software~\cite{esaf}. In case of
Mini-EUSO the overall efficiency of the detector was fine-tuned with ESAF, mainly acting at the level of
MAPMT response, to match the measured one $\epsilon_{ME}$ = 0.080 $\pm$ 0.015 (see the end-to-end in-flight
calibration of Mini-EUSO reported in~\cite{hiroko2}) for
a point-like source on ground. A flat diffused UV emission in the range $\lambda$ = 300 - 400 nm was
simulated at the detector's aperture either with a range of zenith directions much larger than the FoV of
the instrument ($\pm$60$^\circ$ for both detectors) or just within the FoV of the detectors
($\pm$30$^\circ$ for JEM-EUSO and $\pm$22$^\circ$ for Mini-EUSO). The estimated background ratio ($R(ME/JE)$)
between Mini–EUSO and JEM–EUSO at FS level is $R(ME/JE)$ = 0.98 -- 1.04 slightly depending on the range of
zenith angles. This result confirms that the expected photon rate from diffuse sources is similar in
JEM-EUSO and Mini-EUSO instruments.

The energy threshold of Mini-EUSO for point-like sources like UHECRs is
roughly 2 orders of magnitude higher than the original JEM-EUSO one. The right side of Fig.~\ref{fig:fov} shows
the trigger efficiency of Mini-EUSO for ESAF simulated proton-generated EASs of different energies. Events
have been
simulated according to a $\sin(2\cdot \theta)$ dependence on a larger area than the FoV of the instrument on
ground to take into account
border effects. The efficiency ($\epsilon$) as a function of the energy $(E)$ has then been calculated as:
\begin{equation}
\epsilon(E) = \frac{N_{trig}}{N_{simu}} \cdot \frac{A_{simu}}{A_{FoV}} ,
\label{eq:effi}
\end{equation}
where $N_{trig}$ and $N_{simu}$ represent the number of triggered events over the simulated ones for a
specific energy bin, while $A_{simu}$ and $A_{FoV}$ indicate the area on which EAS have been injected and
the area on ground in the FoV of Mini-EUSO, respectively.
The red and violet points assume a fixed nightglow background of 1 and 2 counts/pix/GTU, respectively.
The 50\% trigger efficiency is located between 3 - 5 $\times$ 10$^{21}$ eV depending on the UV nightglow
level.

\begin{figure}[h]
\centering
\includegraphics[width=\columnwidth]{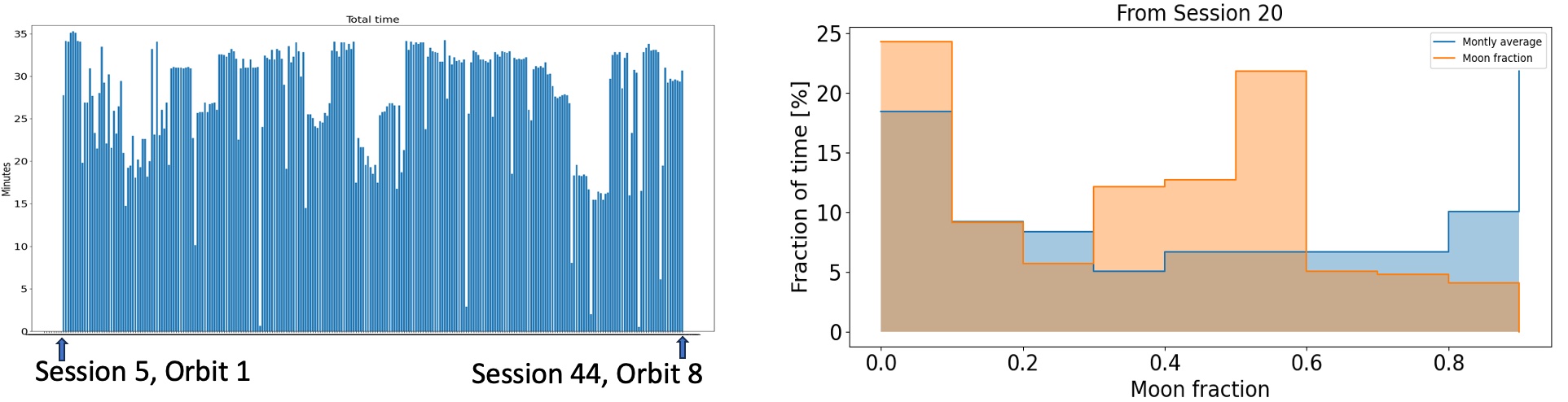}
\caption{Left: Time in minutes in which Mini-EUSO was operational during the various orbits between session 05 and session 44. Right: Brown histogram displays the fraction of time in which Mini-EUSO was operational between sessions 20 and 44 with a specific MB while blue histogram shows the monthly averaged fraction of time in which the Moon appears in the sky in a specific MB.}
%Both histograms show on the X-axis the Moon fraction.}
\label{fig:operations}       % Give a unique label
\end{figure}

The D3 data taken by Mini-EUSO allow a first comparison with the assumed background levels in JEM-EUSO,
K-EUSO and POEMMA to verify that the estimated performance is based on justified assumptions.
The analysis presented in the following is preliminary as it is based on part of the data collected in
25 sessions between session 20 and 44.
A more refined analysis will be performed when all the data collected by Mini-EUSO is available.
The left side of Fig.~\ref{fig:operations} displays the typical number of minutes in which Mini-EUSO was in operation during all data taking sessions between Session 5 and Session 44. The different time of operation in each orbit reflects several aspects such as the different duration of the night in that specific orbit or the moon condition or the switch on and off by the astronauts among others. The right side of Fig.~\ref{fig:operations} displays in brown colour the fraction of time from session 20 to session 44 in which Mini-EUSO operated in a specific Moon Bin (MB). This is compared to the monthly averaged fraction of time in which the Moon appears in the sky with a specific MB (blue colour). Mini-EUSO operated in all the possible Moon conditions with a preference for low Moon phases. This is done on purpose to maximize the scientific potential of Mini-EUSO for atmospheric science (i.e. meteors, elves).

\begin{figure}[h]
\centering
\includegraphics[width=\columnwidth]{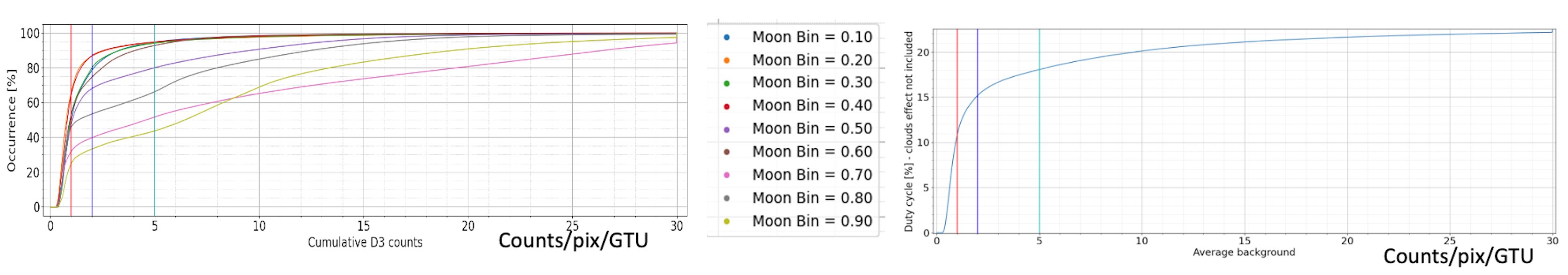}
\caption{Left: Cumulative fraction of time in which Mini-EUSO measured a UV intensity higher than a specific value reported on the X-axis. Different colours refer to the different MB. Each phase has its own normalization. Right: Same as above but summing up all above distribution each weighted for the expected monthly average fraction of time with a specific MB (see text for details).}
%Both histograms show on the X-axis the Moon fraction.}
\label{fig:dutycycle}       % Give a unique label
\end{figure}

According to Mini-EUSO results in low MB conditions, in $\sim$90\% of the time of clear sea conditions
the count rate is below 2.0 counts/pix/GTU, the median being $\sim$0.8 counts/pix/GTU
(see Fig.~\ref{fig:dutycycle}). In clear land conditions there is a higher probability of very low
counts. These are associated with deserts and forests (see~\cite{minieuso-uv} for details). 
Under the reasonable  assumption
that conditions above 2 counts/pix/GTU on land are due to the presence of anthropogenic lights ($\sim$20\% of the time)
and taking into account $\sim$30\% land coverage on Earth, this result corresponds to $\sim$6\% of the total
fraction of time, which is close to the 7\% estimation performed for JEM-EUSO in~\cite{jemeuso}.
Moreover, cloudy conditions typically shift curves by a factor 1.5 - 2 towards higher count rates as
already measured in JEM-EUSO balloon flights~\cite{eusobal}.
Taking into account the fact that JEM-EUSO and Mini-EUSO are expected to measure similar count rates from
diffuse sources, these results are consistent with the assumption done in~\cite{jemeuso} for JEM-EUSO where
the exposure was calculated under the nominal condition of $\sim$1.1 count/pix/GTU.

In order to have a proper estimation of Mini-EUSO exposure at different energies it is necessary to
compute an efficiency curve for the various background levels and convolve each efficiency curve with the
fraction of time in which such background level is measured.
The methodology is described in detail in~\cite{mario-minieuso} and we report here only the essential points. 
As it is not possible to simulate the efficiency
curves for infinite background levels, as a first step the efficiency curves are estimated by simulated proton-generated EASs of various energies only for a limited set of
background levels in the range [0.2;4.0] counts/pix/GTU and the obtained efficiency curves are fitted with a function of the form:
\begin{equation}
\epsilon (E) = 0.5 \cdot [1 + erf(\frac{log_{10}(E/eV) - P_0}{P_1})],
\label{eq:error}
\end{equation}
where the parameter $P_0$ indicates the energy ($E$) at which the efficiency reaches 50\% and $P_1$ measures
the slope of the efficiency curve ($\epsilon(E)$). Subsequently, the obtained parameter values at different
background levels have been fitted with appropriate functions (logarithmic and exponential functions for $P_0$ and $P_1$, respectively) and the dependencies of $P_0$ and $P_1$ as function of the background level have been obtained. Finally, the trigger efficiency curves are modelled at all background levels by equation (\ref{eq:error}) using as $P_0$ and $P_1$ derived by means of the two functions.

Next, the convolved exposure curve ($\mathcal{E}(E)$)
has been calculated according to equation:
\begin{equation}
\mathcal{E}(E) = \sum_i(\epsilon_i(E) \cdot t_i \cdot A_i) \cdot \Omega \cdot t_{tot} \cdot k ,
\label{eq:efficienza}
\end{equation}
where $i$ denotes the UV background level, $t_i$ represents the relative fraction of time in which the
measured background is in the interval $i$ (each interval having a width of 0.01 counts/pix/GTU), $A_i$
accounts for the
cumulative area on ground with background $i$, $\Omega$ measures the UHECR observational solid angle
(which assumes the value of $\pi$),
and $t_{tot}$ = 69.16 hours is the total accumulated time in {\it Cathode3} mode. The parameter $k$ represents the efficiency
factor for detecting
the EAS maximum in presence of clouds. This value depends on the type of clouds, their optical depth and
height. A detailed analysis was performed at the time of JEM-EUSO and adopted in this calculation ($k$
= 72\%).

In order to have a more realistic estimation of the exposure curve, a renormalization has been applied to the fraction of time in which Mini-EUSO has measured a specific UV background level during a specific MB (see left panel of Fig.~\ref{fig:dutycycle}) by rescaling it for the ratio between the fraction of time in which Mini-EUSO operated with a specific MB and the expected monthly average fraction of time according to the Moon cycle (see right panel of Fig.~\ref{fig:operations}). This is summarized in equation:
\begin{equation}
t_i = \sum_{MB} \frac{T_i(MB)}{F(MB)},
\label{eq:tempo}
\end{equation}
where $T_i(MB)$ is the fraction of time in which Mini-EUSO measured the background $i$ in MB and $F(MB)$ is the conversion factor for each MB in order to get the expected monthly averaged fraction of time in a specific MB according to the Moon cycle. 

At the same time the duty cycle of Mini-EUSO as a function of the
tolerated background level ($\eta_i$) can be estimated by the following equation:
\begin{equation}
\eta_i = \frac{1}{\Delta T_{orbit} \cdot N_{orbit}} \cdot \sum_{j=0}^{i} t_j,
\label{eq:duty}
\end{equation}
and it is displayed in the right panel of Fig.~\ref{fig:dutycycle}. As an example it corresponds to $\eta \sim$ 15 (18)\% if up to 2 (5) counts/pix/GTU are tolerated to estimate the duty cycle. 
Finally, the exposure curve obtained by integrating all background levels below 2 (5) counts/pix/GTU has been
renormalized using the exposure obtained at the plateau level of 10$^{22}$ eV. The result, which
represents the convolved trigger efficiency curve, is shown as black (green) line in
Fig.~\ref{fig:fov}. The green curve is very close to the efficiency curve obtained at
the fixed background level of 1 count/pix/GTU, indicating that such a curve provides a reasonable
approximation of
the integrated performance in the various UV background conditions. Moreover, by convolving $\eta$ =18\% with $k$ = 72\% an effective conversion factor between aperture and exposure of 13\% is derived which is consistent with the value obtained in JEM-EUSO computations, confirming the assumptions made in those estimations.
Similarly, these conclusions can be extended also to the estimation of K-EUSO and POEMMA trigger efficiency
curves as they have been derived adopting the very same methodology used for JEM-EUSO.

The accumulated exposure of Mini-EUSO in this analysis amounts to $\sim$1400 Linsley.
A very preliminary extrapolation of this result to all good so far recorded data indicates that
Mini-EUSO has potentially accumulated till now an exposure equivalent to $\sim$3600 Linsley.
%This value is
%comparable to those collected so far by ground-based detectors in hybrid mode~\cite{valerio}.
This value is already about 4\% percent of the exposure collected so far by
ground-based detectors, and equivalent to the one accumulated in the case of hybrid data~\cite{novotny,shin}.

\section{Acknowledgements}
%The authors acknowledge all members of the JEM-EUSO Collaboration, especially the Mini-EUSO team.
This work was supported by the Italian Space Agency through the agreement n. 2020-26-Hh.0, by the French
space agency CNES, and by the National Science Centre in Poland grants
2017/27/B/ST9/02162 and 2020/37/B/ST9/01821. This research has been
supported by the Interdisciplinary Scientific and Educational School of Moscow
University ``Fundamental and Applied Space Research'' and by Russian State
Space Corporation Roscosmos. The article has been prepared based on research
materials collected in the space experiment ``UV atmosphere''. We thank the
Altea-Lidal collaboration for providing the orbital data of the ISS.

\bibliography{my-bib-database}

\newpage
{\Large\bf Full Authors list: The JEM-EUSO Collaboration\\}
%{\scriptsize (author-list as of July 15th, 2023 with reorganized affiliations)} \hspace{0.6cm}
%{\scriptsize (version  \today{} \currenttime{})}
%\vspace*{0.5cm}

\begin{sloppypar}
{\small \noindent
S.~Abe$^{ff}$, 
J.H.~Adams Jr.$^{ld}$, 
D.~Allard$^{cb}$,
P.~Alldredge$^{ld}$,
R.~Aloisio$^{ep}$,
L.~Anchordoqui$^{le}$,
A.~Anzalone$^{ed,eh}$, 
E.~Arnone$^{ek,el}$,
M.~Bagheri$^{lh}$,
B.~Baret$^{cb}$
% ...
}
\end{sloppypar}
\vspace*{.3cm}

%%\newpage
{ \footnotesize
\noindent
% Algeria - 2 institutes
% ...
$^{cb}$ Universit\'e de Paris, CNRS, AstroParticule et Cosmologie, F-75013 Paris, France\\
% ...
$^{ed}$ Istituto Nazionale di Fisica Nucleare - Sezione di Catania, Italy\\
% ...
$^{eh}$ INAF - Istituto di Astrofisica Spaziale e Fisica Cosmica di Palermo, Italy\\
% ...
$^{ek}$ Istituto Nazionale di Fisica Nucleare - Sezione di Torino, Italy\\
$^{el}$ Dipartimento di Fisica, Universit\`a di Torino, Italy\\
% ...
$^{ep}$ Gran Sasso Science Institute, L'Aquila, Italy\\
% Japan - 7 institutes 
% ...
$^{ff}$ Nihon University Chiyoda, Tokyo, Japan\\ 
% ...
$^{ld}$ University of Alabama in Huntsville, Huntsville, AL, USA\\
$^{le}$ Lehman College, City University of New York (CUNY), NY, USA\\
% ...
$^{lh}$ Georgia Institute of Technology, USA\\
% ...
}

%\begin{thebibliography}{99}
%\bibitem{...}
%....

%\end{thebibliography}

%% Full authors list (ONLY FOR COLLABORATIONS)
%\clearpage
%\section*{Full Authors List: \Coll\ Collaboration}
%
%\noindent \textbf{Note comment afterwards:} Collaborations have the possibility to provide an authors list in xml format which will be used while generating the DOI entries making the full authors list searchable in databases like Inspire HEP. \\
%
%\scriptsize
%\noindent
%first.author$^1$, 
%second.author$^2$, 
%third.author$^3$ % .... more names
%and 
%last.author$^{n}$ \\
%
%\noindent
%$^1$first.affiliation.
%$^2$second.affiliation. % .... more affiliation
%$^{m}$last.affiliation.

\end{document}